\documentclass[10pt]{article}

\usepackage[preprint]{tmlr}

\usepackage{amsmath,amsfonts,bm}

\def\eqref#1{equation~\ref{#1}}

\def\1{\bm{1}}

\DeclareMathAlphabet{\mathsfit}{\encodingdefault}{\sfdefault}{m}{sl}
\SetMathAlphabet{\mathsfit}{bold}{\encodingdefault}{\sfdefault}{bx}{n}

\usepackage{hyperref}
\usepackage{url}
\usepackage{graphicx, color}

\usepackage[noorphans]{quoting}
\usepackage[square,sort,comma,numbers]{natbib}

\usepackage{amsmath}
\usepackage{amssymb}
\usepackage{mathtools}
\usepackage{amsthm}
\usepackage{bbm}
\usepackage{booktabs}
\usepackage{tabularx}
\usepackage{subcaption}
\usepackage{wrapfig}

\usepackage{caption}

\usepackage[most]{tcolorbox}

\usepackage[ruled, algo2e, vlined]{algorithm2e}
\SetKwInput{KwInput}{Input}
\SetKwInput{KwOutput}{Output}

\theoremstyle{plain}

\theoremstyle{definition}

\theoremstyle{remark}

\allowdisplaybreaks

\usepackage{listings}
\usepackage{xcolor}
\definecolor{codegreen}{rgb}{0,0.6,0}
\definecolor{codegray}{rgb}{0.5,0.5,0.5}
\definecolor{codepurple}{rgb}{0.58,0,0.82}
\definecolor{backcolour}{rgb}{1,1,1}

\lstdefinestyle{mystyle}{
    backgroundcolor=\color{backcolour},   
    commentstyle=\color{codegreen},
    keywordstyle=\color{magenta},
    numberstyle=\tiny\color{codegray},
    stringstyle=\color{codepurple},
    basicstyle=\ttfamily\footnotesize, 
    breakatwhitespace=false,         
    breaklines=true,                 
    captionpos=b,                    
    keepspaces=true,                 
    numbers=left,                    
    numbersep=5pt,                  
    showspaces=false,                
    showstringspaces=false,
    showtabs=false,                  
    tabsize=4,
    frame=lines      
}

\title{AgentRecommender: LLM Agents Enable Customizable Recommender Systems on the User Side}

\author{\name Ryoma Sato \email rsato@nii.ac.jp \\
  \addr National Institute of Informatics
}

\begin{document}

\maketitle

\begin{abstract}
Recommender systems have traditionally been developed for platforms. However, this has given rise to many phenomena that may be advantageous for platform lock-in but are a nuisance to users, such as clickbait, filter bubbles, and the spread of fake news. Recently, user-side recommender systems have been proposed as a new paradigm for solving this problem. If users deploy their own recommender systems, they are no longer at the mercy of the platform's interests. However, building a user-side recommender system is not trivial; in particular, customizing one for oneself requires additional data. We propose AgentRecommender, a method that leverages the investigation capability and internal knowledge of LLM agents to flexibly build user-side recommender systems without additional data. AgentRecommender allows users to easily create recommender systems tailored to their own preferences.
\end{abstract}

\section{Introduction}

Since their success at Amazon \cite{linden2003amazon} and Netflix \cite{gomez2016netflix}, recommender systems have been developed for platforms. To increase sales, they recommend products by predicting the probability that a user will purchase them; to keep users on the platform longer, or to make them watch more advertisements, they recommend content by predicting expected watch time. Recommender systems have brought platforms substantial profits. It is said that 35 percent of the sales at Amazon \cite{mckinsey2013retail}, 80 percent of the time consumed at Netflix \cite{gomez2016netflix}, and more than 50 percent of the content on Instagram \cite{deng2024meta} are delivered through recommender systems.

It is true that recommender systems have helped platforms grow, and users have also often been helped by recommender systems in finding items they like. However, recommender systems have not always worked purely in users' favor. A recommender system that optimizes the probability of purchase may recommend products that merely look good and do not serve the user, and in the hands of a recommender system that maximizes expected watch time, the user may end up being shown one video after another that keeps dragging out its conclusion. Recommender systems have also caused more serious problems, such as filter bubbles \cite{pariser2011filter,hussein2020measuring,youtube2020conspiracy,tomlein2021audit} and unfairness \cite{hardt2016equality, kamishima2012fairness, zafar2017fairness}. Platform operators may fix these problems immediately when they hurt their own interests, for example when users start leaving, but otherwise they may not actively fix them. In such cases, there is little that users can do.

To solve this problem, \citet{sato2022private} proposed user-side recommender systems, which allow users to build recommender systems with the functionality they desire by themselves. For example, when the platform's recommender system delivers news biased toward a particular political party, a user-side recommender system can balance the parties of the recommended items. When the platform's recommender system delivers only popular items, a user-side recommender system can recommend minor yet relevant items. Unlike the platform, users are under severe constraints: they cannot directly access the database, nor can they obtain the log data of other users. \citet{sato2022private} solved this problem by cleverly using the recommendation network.

However, conventional user-side recommender systems \cite{sato2022private,sato2022towards,sato2024overhead} require item labels to implement the desired functionality. For example, when correcting for popularity, play counts or like counts can be used directly if they are public, as on video streaming services, but a movie streaming site may not disclose view counts. Likewise, some news sites may provide tags, but when they do not, telling which political party an article favors is not trivial. One could resolve this by using an additional classifier, but preparing a classifier from scratch is a heavy burden for an ordinary user.

We solve this problem by using LLM agents. Since LLM agents possess world knowledge, they can often estimate item attributes zero-shot, and when necessary, they can estimate item attributes even more reliably by combining this with search and other tools. By combining them with user-side recommender systems, users can easily build recommender systems with the desired functionality. The proposed method, AgentRecommender, can realize the desired functionality on the user's side even in regimes where conventional user-side recommender systems do not work.

\section{Problem Setting}

Following the setting of \citet{sato2022private}, we assume that the service provider offers an item-to-item recommender system. That is, when the page of item $i \in \mathcal{I}$ is accessed, a set of related items $\mathcal{P}_{\text{Prov}}(i) \in \mathcal{I}^K$ is provided, where $\mathcal{I}$ is the set of all items. The service provider's recommender system is not necessarily fair, nor does it necessarily have the properties the user desires. As such a system, we can exploit, for example, Amazon's ``Customers Who Bought This Item Also Bought'' section or YouTube's "Up next" section. We also assume that each item $i$ has a hidden attribute $a_i \in \mathcal{A}$. Any kind of attribute is conceivable: for example, genres such as $\mathcal{A} = \{\text{``action''}, \text{``horror''}, \text{``comedy''}\}$, eras such as $\mathcal{A} = \{\text{``1990s''}, \text{``2000s''}, \text{``2010s''}, \text{``2020s''}\}$, popularity such as $\mathcal{A} = \{\text{``major''}, \text{``minor''}\}$, price ranges such as $\mathcal{A} = \{\text{``low-price''}, \text{``high-price''}\}$, and political leanings such as $\mathcal{A} = \{\text{``Democratic''}, \text{``Republican''}\}$. The user may design this attribute partition arbitrarily; we make no assumptions whatsoever about the categorization and leave it to the user. This is precisely the strength of realizing arbitrary functionality with user-side recommender systems. We assume that the user wants the recommended items to have these attributes in specific proportions, to be balanced, or to achieve higher coverage in the sense that at least one item from each category is recommended. The service provider's recommender system does not necessarily respect these constraints. Under such circumstances, the user builds a recommender system that respects the desired constraints. A user who is tired of being recommended only famous items may specify popularity as the category so as to receive recommendations with balanced popularity, and a user who is fed up with recommendations biased toward a particular political leaning may specify political leaning as the category so as to receive politically balanced recommendations. The major difference from previous user-side recommender systems is that we assume the attribute $a_i \in \mathcal{A}$ is not observable to the user, in contrast to what previous user-side recommender systems assumed. This drastically reduces the amount of data the user needs to build a recommender system. The attributes $a_i \in \mathcal{A}$ are observed only at evaluation time, where we evaluate how well the recommender system built by the user satisfies the constraints.

In summary, the problem setting is stated as follows.

\begin{tcolorbox}[colframe=gray!20,colback=gray!20,sharp corners]
    \textbf{Given:} Oracle access to the official recommendations $\mathcal{P}_{\text{prov}}$; the attribute space $\mathcal{A}$ and the constraints to be satisfied, specified by the user in natural language; and a budget $B$. The attribute $a_i \in \mathcal{A}$ of each item is not observed at construction time and is referred to only at evaluation time. \\
    \textbf{Output:} A user-side recommender system $\mathcal{Q}\colon \mathcal{I} \to \mathcal{I}^K$ that satisfies the constraints on $\mathcal{A}$.\\
    Internally, $\mathcal{Q}$ may evaluate $\mathcal{P}_{\text{prov}}$ at most $B$ times per recommendation.
\end{tcolorbox}

For simplicity, we consider user-side item-to-item recommendation here, but user-side user-to-item recommendation can be constructed in the same manner.

\section{Proposed Method: AgentRecommender}

\begin{figure}[t]
  \centering
  \includegraphics[width=0.8\textwidth]{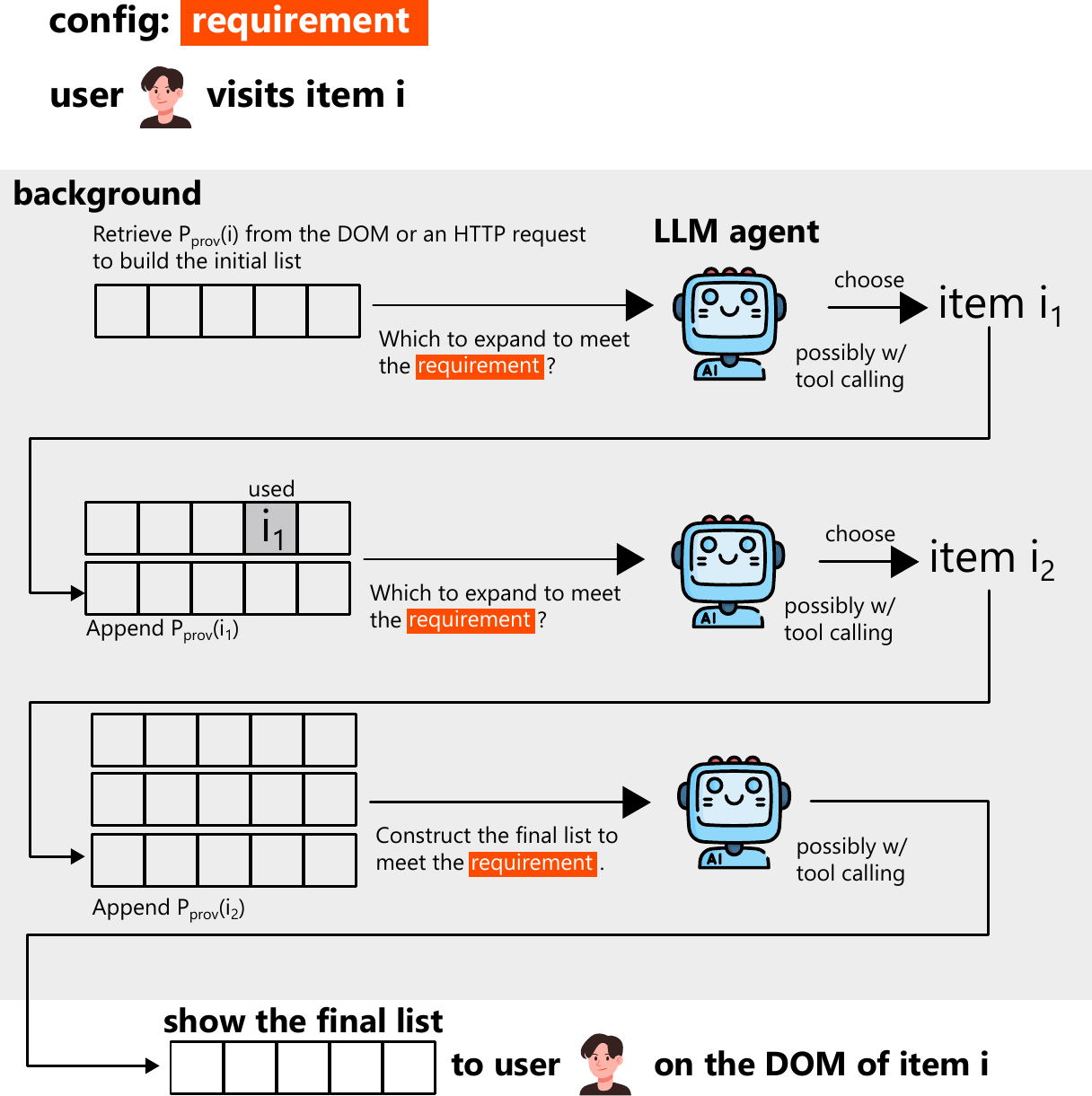}%
  \caption{Conceptual illustration of AgentRecommender. AgentRecommender uses an LLM agent to build a recommendation list that satisfies the constraints while using the service provider's recommender system $\mathcal{P}_{\text{prov}}$.} \label{fig: illust}
\end{figure}

The proposed method, AgentRecommender, uses an LLM agent to build a recommendation list that satisfies the constraints on the item attributes $\mathcal{A}$ while using the service provider's recommender system $\mathcal{P}_{\text{prov}}$ (Figure \ref{fig: illust}). AgentRecommender repeatedly observes $\mathcal{P}_{\text{prov}}(j)$ of an item $j$ that the LLM deems promising and adds it to the candidates, thereby gradually collecting promising items, and in the end, the LLM constructs a list that satisfies the conditions from the candidates. This process can also be viewed as an LLM agent actively exploring the recommendation network, i.e., the graph whose nodes are the items and in which an edge is drawn from $i$ to $j$ when item $j$ is recommended on the page of item $i$.

Suppose the user accesses the page of item $i$. The service provider's recommendations $L = \mathcal{P}_{\text{prov}}(i) \in \mathcal{I}^K$ can then be observed from the DOM, HTTP requests, and the like. Starting from this list as the initial list, we repeat the following operations $B$ times using the LLM: \begin{enumerate}
  \item Present to the LLM, in natural language, the items in the current list that have not yet been expanded together with the requirements, and let it select the item to expand.
  \item Access the page of the selected item $j$ behind the browser, obtain the service provider's recommendations $\mathcal{P}_{\text{prov}}(j) \in \mathcal{I}^K$, and add them to the list $L$.
\end{enumerate}
Finally, we present the items of $L$ obtained so far and the requirements to the LLM, and let it select the $K$ items that should ultimately be presented to the user.

Only the item names are given to the LLM, but the LLM estimates the attributes of the items using its internal knowledge (and, if necessary, tool calls such as search) and selects the item that should be explored to satisfy the requirements. For example, if the current list is overly skewed toward action movies, the LLM would select a horror movie as the item $j$ to expand. The recommendations for $j$ are likely to contain horror movies, and since they are added to the list, the requirements become easier to satisfy. The list $L$ ends up containing items different from those presented by the original recommender system, but since the included items are at most $B$ hops away from the starting point, they are expected to be highly related to the original recommendations. Moreover, when we finally let the LLM select items from the list, the relevance issue can also be resolved by including in the prompt the requirement that the selection be restricted to items that are sufficiently related to the original item.

We have considered item-to-item recommendation here, but user-side user-to-item recommendation is also possible by using the service provider's user-to-item recommendation results as the initial list $L$.

Issuing $B$ additional queries every time the user accesses an item page may incur too much latency or place a burden on the service. In that case, as in RecCycle \cite{sato2024overhead}, we can exploit a cache of the recommendation results obtained during the user's ordinary use of the service: instead of issuing queries online, we issue queries to a cache database built in, for example, the browser's localstorage, thereby suppressing both the latency and the load on the service.

\section{Experiments}

We validate the effectiveness of AgentRecommender on three datasets: MovieLens 1M (movies) \cite{harper2016movielens}, LastFM (music) \cite{cantador2011second}, and Amazon Home and Kitchens (e-commerce) \cite{he2016ups,mcauley2015image}. As the LLM, we use GPT-5.6 Luna (reasoning\_effort=low) and receive its outputs as structured outputs with JSON schemas. In each expansion turn, we constrain the output with an enum listing the candidate ids and let the LLM select exactly one item; in the final selection, we let it select $K$ distinct items from the accumulated candidate pool.

The experimental setup is as follows. First, we fix a single user. We perform item-to-item recommendation for the item this user interacted with last. The service provider's recommender $\mathcal{P}_{\text{prov}}$ is implemented by obtaining item embeddings with Bayesian Personalized Ranking (BPR) \cite{rendle2009bpr} and recommending the items whose embeddings are close. Note that neither the embeddings nor the fact that the system is built with BPR can be exploited when building the user-side recommender system: at construction time, only black-box access to $\mathcal{P}_{\text{prov}}$ is allowed. As the requirements on the categories, we use one requiring that every category be included in the recommendation results at least $\tau$ times, for which we measure the rate at which it is achieved, and one requiring that the recommendation results include as many kinds of categories as possible, for which we measure the average number of unique categories. Throughout the following experiments, we set the number of recommendations to $K = 10$ and the budget to $B = 8$. We randomly select $n = 100$ users from the dataset and build a recommender system on the user's side. We report the mean and standard deviation of the evaluation values across users.

\begin{table}[t]
\centering
\caption{\textbf{Experimental Results}. The proposed method, AgentRecommender, achieves far higher performance than the service provider's recommender system and than merely re-ranking it, across the three domains of MovieLens 1M (movies), LastFM (music), and Amazon Home and Kitchens (e-commerce) and the three criteria of oldness, genre, and price. This shows that even when the service provider's recommender system does not offer a particular functionality, users can build a recommender system with the desired functionality on their own. }
\label{tab: results}
\begin{tabular}{lcccc}
\toprule
& ML-1M & ML-1M & LastFM & Amazon \\
& success (oldness) $\uparrow$ & coverage (genre) $\uparrow$ & coverage (genre) $\uparrow$ & success (price) $\uparrow$ \\
\midrule
Provider & $0.00 \pm 0.00$ & $5.51 \pm 2.35$ & $2.50 \pm 1.02$ & $0.16 \pm 0.37$ \\
iAgent \cite{xu2025instructagent}  & $0.00 \pm 0.00$ & $5.51 \pm 2.35$ & $2.50 \pm 1.02$ & $0.16 \pm 0.37$ \\
AgentRecommender & $\mathbf{0.68 \pm 0.47}$ & $\mathbf{8.94 \pm 2.43}$ & $\mathbf{4.34 \pm 1.62}$ & $\mathbf{0.64 \pm 0.48}$ \\
\bottomrule
\end{tabular}
\end{table}

\subsection{MovieLens 1M}

MovieLens 1M is a movie rating dataset containing 3,883 movies. We conduct experiments with the goal of being recommended movies of various genres and eras. First, using genre as the category, we evaluate the number of distinct genres included among the $K = 10$ recommended items; larger values indicate more diversity and are better. Note that a single movie can belong to multiple genres. Second, we set $a_i = \text{old}$ if the movie was released in 1989 or earlier and $a_i = \text{new}$ if it was released in 1990 or later, and we evaluate the rate at which both are included in the recommendation results with $\tau = 5$ items each. Note that the release year at the end of a movie title (e.g., ``(1977)'' in ``Star Wars (1977)'') is removed from the titles presented to the agent. Therefore, the LLM cannot read the year directly from the title and must estimate oldness based on world knowledge. Larger values indicate a higher rate of balanced recommendations and are better.

The results are reported in the first and second columns of Table \ref{tab: results}. The provider and iAgent do not succeed at all under the oldness criterion. This is because many recommender systems recommend only new items for new items and only old items for old items. However, this makes the recommendations biased, and users may be dissatisfied. Since iAgent merely reorders the items, it cannot improve the score over the provider's recommender system. Even under such a difficult criterion, the proposed AgentRecommender produces recommendations with spread-out release years in as many as 68 percent of the cases. In the remaining 32 percent, it does not fail completely either: although the categories are not split half-and-half, the recommendations come close to it. Under the genre coverage criterion as well, AgentRecommender succeeds in including, on average, 1.6 times as many genres in the recommendation list as the provider and iAgent. These results show that AgentRecommender is beneficial for users who want diverse recommendations.

\subsection{LastFM}

LastFM is a dataset of music artists. We filter out artists with fewer than 15 users, leaving 1,052 artists after preprocessing. We conduct experiments with the goal of being recommended artists of various genres. The evaluation value is the number of distinct genres included among the $K = 10$ recommended items; larger values indicate more diversity and are better.

The results are reported in the third column of Table \ref{tab: results}. Here as well, as with MovieLens, AgentRecommender succeeds in including, on average, 1.7 times as many genres in the recommendation list as the provider and iAgent. These results show that AgentRecommender is beneficial for users who want diverse recommendations.

\begin{figure}[t]
  \centering
  \includegraphics[width=0.8\textwidth]{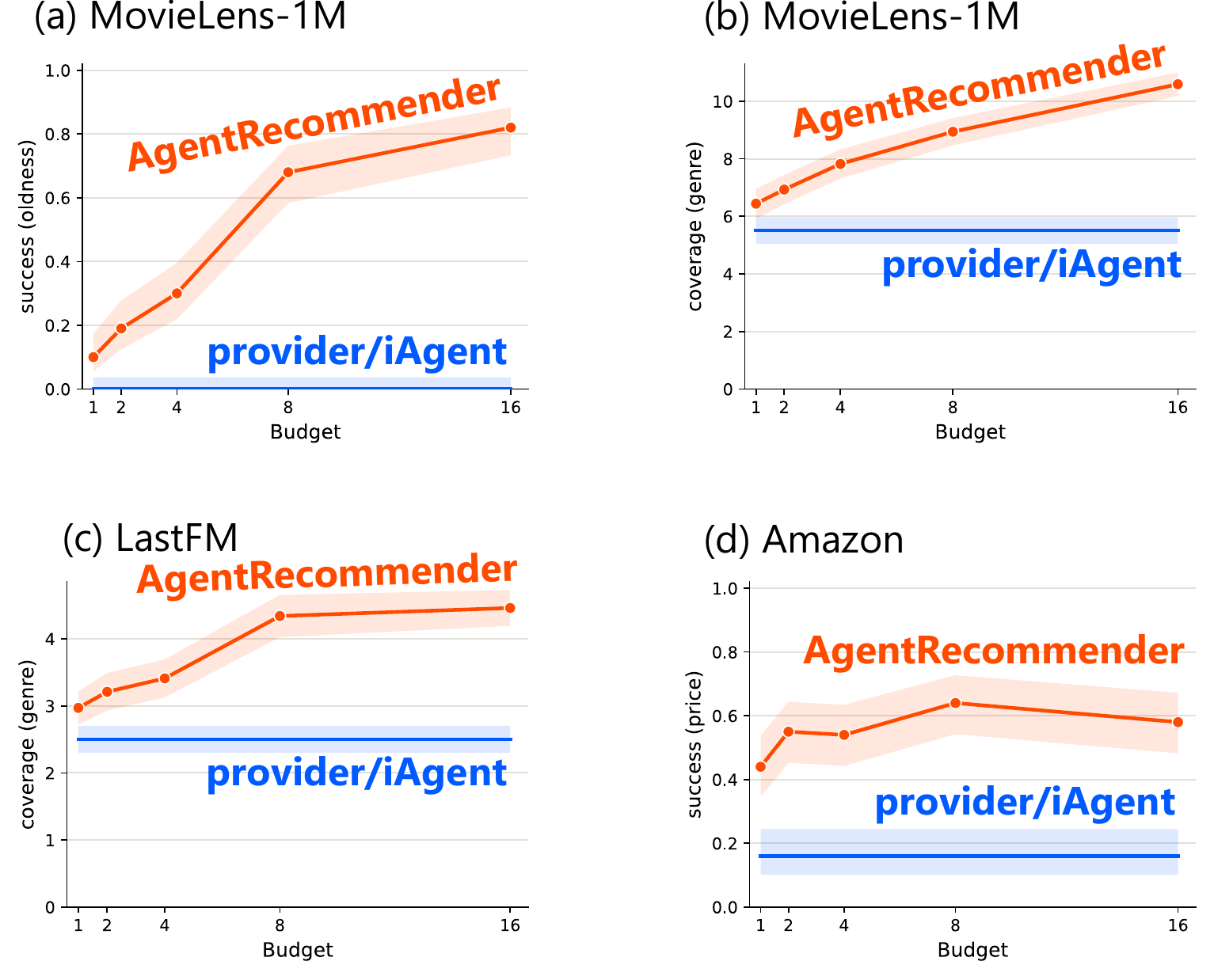}%
  \caption{Budget variation experiment. AgentRecommender greatly outperforms the provider and iAgent even with a budget of $B = 1$, and it can improve its performance further as the budget increases. The bands represent 95 percent confidence intervals.} \label{fig: budget}
\end{figure}

\subsection{Amazon Home and Kitchens}

Amazon Home and Kitchens is a product review dataset. We start from the 5-core subset released by McAuley et al. (every user and every item has at least five reviews), and from it we keep only the products that have a title, have a price of at least \$1, and have a Home and Kitchen subcategory (the second level of the category path, such as Kitchen \& Dining or Bedding). After this filtering, we keep only the users who have interacted with at least two of the remaining items. After preprocessing, 23,927 products remain. Among e-commerce recommender systems, one that recommends simply based on relatedness or co-purchases may recommend only low-priced products for low-priced products, and the user's product search may hit a dead end; a recommender system that aims to maximize the platform's profit may aggressively recommend only high-priced products, and the user may end up buying an expensive product without noticing the existence of cheaper products that would actually be sufficient. We conduct experiments with the goal of being recommended both low-priced and high-priced products. Specifically, we take the median price of the products in this category, \$19.99, as the threshold, and set $a_i = \text{low-price}$ for products below \$19.99 and $a_i = \text{high-price}$ for products at or above \$19.99. We evaluate the rate at which both are included in the recommendation results with $\tau = 5$ items each. Larger values indicate a higher rate of balanced recommendations and are better.

The results are reported in the fourth column of Table \ref{tab: results}. Here as well, AgentRecommender achieves, on average, four times the success rate of the provider and iAgent. In other words, for users who want to consider both cheap and expensive products, AgentRecommender can provide such opportunities even when the service provider's official recommender system does not.

\subsection{Budget Sensitivity}

So far, we have fixed the budget to $B = 8$. Figure \ref{fig: budget} shows the results when the budget is varied from $B = 1$ to $B = 16$. In all four settings, AgentRecommender already greatly outperforms the provider and iAgent at $B = 1$. This shows that AgentRecommender works even with a small budget and is practical. As the budget increases, the performance generally improves. The exception is the Amazon dataset; we attribute this to the fact that, on the Amazon dataset, estimating prices from product names alone is difficult, so even when a larger budget provides an abundant pool of candidate products, achieving a perfect price balance remains difficult, and the success rate plateaus around 0.6.

\section{Related Work}

\textbf{User-side Realization.} User-side realization refers to realizing desired functionality of a service on the user's side \cite{sato2024user}. Even a service that a user finds convenient and keeps using is bound to have some inconveniences. However, users have little power, and sending requests to the service does not guarantee that the inconvenience will be resolved. Since the data and content are locked in on the platform side, it is difficult for users to build a service from scratch by themselves. In this situation, user-side realization algorithms take a middle path: they realize the functionality the user desires while still using the service. Representative examples include user-side search engines \cite{sato2022retrieving,sato2022clear}, user-side privacy protection \cite{sato2024making}, and user-side watermarking of large language models \cite{sato2023embarrassingly}.

Most closely related to this work are user-side recommender systems. PrivateWalk and PrivateRank are the first user-side recommender systems \cite{sato2022private}.  However, they are not practical because they require a large number of accesses to the official recommender system or degrade the performance. Consul \cite{sato2022towards} is the first practical user-side recommender system with multiple desirable properties. RecCycle \cite{sato2024overhead} improved the efficiency further and realized an effective user-side recommender system with no overhead. However, all of these require item labels, which becomes a burden on the user when the service does not provide them. iAgent \cite{xu2025instructagent} is a recently proposed user-side recommender system that uses an agent. However, it merely reorders the items presented by the service, and it essentially cannot discover new items. The proposed method is the first user-side recommender system that requires no labels and can discover new, useful items. As a closely related idea, Bonsai \cite{malki2026bonsai} personalizes social feeds by describing the user's preferences in natural language and selecting and ranking posts while performing search and other operations. Alexandria \cite{kolluri2026alexandria} and Value Alignment \cite{jahanbakhsh2026value} are also user-side methods for personalizing social feeds: when the user specifies preferred attributes, an LLM predicts the attributes of the feed and reorders the feed items to match the user's preferences. These can also be regarded as a kind of user-side realization. \citet{stray2026prosocial} conducted a large-scale examination of the effects of such user-side re-ranking of social feeds. \citet{yuan2026recommender} and \citet{lazar2024moral} argue for making recommender systems agent-centric, partly for the sake of user-side controllability. This argument is also made in \citet{sato2024user}. In a parallel direction, there is also a method that transfers recommendation data locked into one service to another service on the user's side \cite{sato2025solving}, which can be used as a direction orthogonal to the present method.

Separately from these lines of work, methods that incorporate LLM agents into recommender systems have also been proposed \cite{yang2026agentdr,zhang2026recthinker,maragheh2025arag,yu2026thought,kong2025think,raj2026harpo,huang2025mr,wang2025tunable,yang2026personalized,khezresmaeilzadeh2025preserving,wozniak2025improving}. However, these are basically assumed to run on the service provider's servers, not on the user's side. For example, TEARS \cite{penaloza2025tears} enables users to convey their preferences to the recommender system in natural language, but it runs on the service provider's servers, so users cannot use it unless the service provider implements TEARS. The advantage of user-side realization is that it can be achieved by the user's own power alone, without the provider implementing the method or offering endpoints. In addition, methods that simulate users with LLMs, such as AgentCF \cite{zhang2024agentcf,liu2025agentcf} and RecAgent \cite{wang2024user}, have been proposed, but they are mainly used for modeling and evaluation on the provider side.

\vspace{0.1in}
\noindent \textbf{Fairness in Recommender Systems.} As fairness has become a major concern in society \cite{united2014big, executive2016big}, many fairness-aware machine learning algorithms have been proposed \cite{hardt2016equality, kamishima2012fairness, zafar2017fairness}. In particular, fairness with respect to gender \cite{zehlike2017fair, singh2018fairness, xu2020algorithmic}, race \cite{zehlike2017fair, xu2020algorithmic}, financial status \cite{fu2020fairness}, and popularity \cite{mehrotra2018towards, xiao2019beyond} is of great concern. In light of this, many fairness-aware recommendation algorithms have been proposed \cite{kamishima2012enhancement, yao2017beyond, biega2018equity, milano2020recommender}. Some of them aim to ensure fairness for users \cite{bruke2017mltisided} and others aim to ensure fairness for items (such as products and accounts) \cite{ekstrand2018exploring, beutel2019fairness, mehrotra2018towards, liu2019personalized, geyik2018building}, and some aim to ensure fairness for both users and items \cite{bruke2017mltisided}. In this study, we focus on fairness for items following \cite{sato2022private, sato2022towards}, but the proposed method can also be applied to fairness for users when the official system is not fair for the very user that uses the user-side system. The user can overcome the unfairness of the official system by creating their own recommender system. Note that fairness is closely related to topic diversification \cite{ziegler2005improving} by regarding the topic as the sensitive attribute, and we considered the diversity of recommended items in this study as well.

\vspace{0.1in}
\noindent \textbf{Steerable Recommender Systems.} The reliability of recommender systems has attracted a lot of attention \cite{tintarev2007survey, balog2019transparent}, and steerable recommender systems that let the users modify the behavior of the system have been proposed \cite{green2009generating, balog2019transparent}. User-side recommender systems also allow the users to modify the recommendation results. However, the crucial difference between steerable and user-side recommender systems is that steerable recommender systems must be implemented by a service provider, whereas user-side recommender systems can be built by arbitrary users even if the official system is an ordinary (non-steerable) one. Therefore, user-side recommender systems can expand the scope of steerable recommender systems by a considerable margin \cite{sato2022private}.

\section{Conclusion}

We proposed AgentRecommender, a method for building a recommender system with desired functionality on the user's side, and moreover without additional labels. AgentRecommender actively explores items that satisfy the desired requirements while using the service provider's recommender system, and presents them to the user. AgentRecommender can realize the desired functionality within the scope of the user's privileges, skillfully steering clear of the areas an ordinary user cannot access, such as the database and the log data of other users. Moreover, by exploiting the internal knowledge and search capability of LLM agents, it avoids burdensome work such as labeling. This makes recommender systems much easier for ordinary users to build. In our experiments, across the three domains of movies, music, and e-commerce, we succeeded in creating recommender systems that satisfy the requirements even though the service providers' recommender systems did not satisfy them.

\bibliography{main}
\bibliographystyle{abbrvnat}

\end{document}